\documentclass[11pt]{article}
\usepackage{graphicx,amsmath,bm, amsthm,mathrsfs,amssymb, braket, verbatim}
\usepackage[usenames]{color}
\usepackage{ulem}
\usepackage{authblk}
\usepackage{cite}
\usepackage{tcolorbox}

\newcommand{\ee}{\end{equation}}

\newcommand{\reff}[1]{(\ref{#1})}
\newcommand{\beq}{\begin{equation}}
\newcommand{\eeq}[1]{\label{#1}\end{equation}}
\newcommand{\beqa}{\begin{eqnarray}}
\newcommand{\eea}{\end{eqnarray}}
\newcommand{\eeqa}[1]{\label{#1}\end{eqnarray}}
\newcommand{\beg}{\begin{equation*}}
\newcommand{\eeg}{\end{equation*}}

\newcommand{\m}{\!-\!}

\newcommand{\bsplit}{\begin{split}}
\newcommand{\esplit}{\end{split}}

\usepackage{subfig} %To put many figures side by side
\usepackage{circuitikz} %To draw the electrical circuits
\usepackage[capposition=bottom]{floatrow} %To add notes under a figure
\usepackage{epigraph} %For quotes at the beginning of sections and text
\usepackage{appendix}
\usepackage{rotating}
\allowdisplaybreaks

\title{New degeneracies and modification of Landau levels in the presence of a parallel linear electric field}

\author[]{Ariel Edery\thanks{aedery@ubishops.ca}}
\author[]{Yann Audin\thanks{yaudin13@ubishops.ca}}

\affil[]{Department of Physics and Astronomy, Bishop's University, 2600 College Street, Sherbrooke, Qu\'{e}bec, Canada, J1M 1Z7.\vspace{1em}}

\begin{document}
\date{}
\maketitle
\begin{abstract}
We consider a three-dimensional system where an electron moves under a constant magnetic field (in the z-direction) and a \textit{linear} electric field parallel to the magnetic field above the z=0 plane and anti-parallel below the plane. The linear electric field leads to harmonic oscillations along the z-direction. There are therefore two frequencies characterizing the system: the usual cyclotron frequency $\omega_c$ corresponding to motion along the x-y plane and associated with Landau levels and a second frequency $\omega_z$ for motion along the z-direction.  Most importantly, when the ratio $W\!\!=\omega_c/\omega_z$ is a rational number, the degeneracy of the energy levels does not remain always constant as the energy increases. At some energies, the degeneracy jumps i.e. it increases. In particular, when the two frequencies are equal, the degeneracy increases with each energy level. This is in stark contrast to the usual Landau levels where the degeneracy is independent of the energy. We derive compact analytical formulas for the degeneracy. We also obtain an analytical formula for the energy levels and plot them as a function of $W$. The increase in degeneracy can readily be seen in the plot at points where lines intersect. For concreteness, we consider the electric field produced by a uniformly charged ring. Besides a linear electric field in the z direction the ring produces an extra electric field in the xy plane which we treat via perturbation theory. The Landau degeneracy is now lifted and replaced by tightly spaced levels that come in ``bands". The plot of the energy levels shows that there is still a degeneracy where the bands intersect.            
\end{abstract}
%\thispagestyle{empty}
%\end{titlepage}
\setcounter{page}{1}
\newpage
\section{Introduction}\label{Intro}
An electron moving in two dimensions (x-y plane) under a constant magnetic field (in the z-direction) has discrete energies known as Landau levels. The energies, neglecting spin, are simply those of the one-dimensional harmonic oscillator with cyclotron frequency $\omega_c$. Authors have now investigated Landau quantization in many different and interesting contexts. It has been studied in the presence of topological defects \cite{Marque}, in the spinning cosmic string spacetime \cite{Muniz}, in the presence of disclinations \cite{Furtado}, under non-inertial and gravitational effects \cite{Konno}, in molecules under the coupling between cyclotron motion and core rotation \cite{Kimura} and relativistically at finite temperature \cite{Bene}. 

In this paper we consider adding a \textit{linear} electric field to the Landau scenario of a constant magnetic field. The electric field is parallel to the magnetic field above the z=0 plane and anti-parallel to the magnetic field below the z=0 plane. This particular system has not been studied before and is of interest for a few reasons. First, the system has two fundamental frequencies and there is an extra degeneracy, beyond the Landau degeneracy, that arises from this. Second, a uniformly charged ring can produce a linear electric field in the $z$ direction in the vicinity of its center though an extra electric field along the x-y plane is also produced and can be treated in perturbation theory. Before discussing this in more detail it is worth noting that the literature contains considerable studies of the effect of non-uniform electric fields on ``Landau-type" systems containing a magnetic field. We review this below.

In \cite{Ref1} an atomic analogue of Landau quantization based on the Aharonov-Casher interaction \cite{Ref11} is developed where a non-uniform electric field is applied in the x-y plane (with magnetic field in the $z$ direction). In \cite{Ref2} the authors study the Landau quantization of neutral particles in an external field. They consider crossed magnetic and electric fields where the electric field is non-uniform and is again along the x-y plane. This field configuration is interesting because it confines the dipole in a plane and produces a coupling similar to the coupling of a charged particle in the presence of external magnetic field (i.e. the analog of Landau levels). Relativistic and curved spacetime landau quantization of a neutral particle with a permanent magnetic dipole moment was then later studied in \cite{Ref3} and \cite{Ref4} respectively where the electric field was again non-uniform and confined to the x-y plane. Authors have also studied in the context of non-commutative quantum mechanics \cite{Ref5} the analog of Landau levels in the presence of electric and magnetic fields (again the electric field was applied in the x-y plane).  Quantum ring quantization was studied for a neutral particle in \cite{Ref6} for two different non-uniform electric field configurations in the x-y plane. In \cite{Ref7} the analog of Landau quantization for neutral atoms with an induced electric dipole was investigated in the presence of topological defects. They show that the infinite degeneracy of the Landau levels is lifted (i.e. broken). Landau-like states in neutral particles were further and recently studied in \cite{Ref8} and with a Kratzer potential in a rotating frame in \cite{Ref9}. Rotating effects on the Landau quantization for an atom with a magnetic quadrupole moment have also been recently studied in \cite{Fonseca1,Fonseca2}.

In the above literature the non-uniform electric field is applied in the x-y plane and typically on a neutral particle or atom. In this paper a linear electric field is applied in the $z$ direction on a charged particle. This leads the particle to oscillate harmonically about z=0 along the z-direction with frequency $\omega_z$. We therefore obtain a three-dimensional system with two characteristic frequencies: $\omega_c$ and $\omega_z$. The linear electric field clearly modifies the energy levels of the electron. But the most crucial point is that the degeneracy of the system does not necessarily remain constant as the energy increases when the ratio of the two frequencies, $W\!=\omega_c/\omega_z$, is a rational number. At particular energies, the degeneracy will jump i.e. it will increase. This is most apparent when the two frequencies are equal. In that case, the degeneracy increases with each energy level. Recall that the Landau levels themselves have a significant degeneracy (referred to in this paper as the Landau degeneracy) but it is constant for every energy level. Adding the linear electric field therefore alters the degeneracy of the system in a profound way. We calculate the degeneracy and obtain compact analytical formulas for them. We also obtain analytical formulas for the energy levels as a function of $W$ and two quantum numbers $n$ and $n_z$. We plot the energy as a function of $W$ and each line represents a pair $(n,n_z)$. The degeneracies increase at points where lines for different $(n,n_z)$ pairs intersect. The increase in degeneracy is reminiscent of the two-dimensional Fock-Darwin system \cite{Fock,Darwin} where besides an applied constant magnetic field, an extra quadratic potential of the form $\frac{1}{2} m \,\omega_0^2\,(x^2+y^2)$ is often added by hand in order to confine the electrons in the x-y plane. Such a system is characterized by two frequencies, $\omega_c$ and $\omega_0$, and the degeneracy is also altered significantly for particular ratios of the two frequencies. The Fock-Darwin system has been very popular and successful in studying quantum dots \cite{Madhav,Burkard,Kuan,Babinski,Maharovsky,Efros,Avetisyan}, a major area of research in nanotechnology. It should be noted that other types of confinement, this time along the z direction, may be found in nanoelectronic devices. In particular, one can consider a linear (triangular quantum well) potential (e.g. see \cite{Arora}). The solution along the z direction would then be given by Airy functions and this would also lead to a higher density of levels with energy. 

A linear electric field in the $z$ direction can be produced in the vicinity of the center of a uniformly charged ring. However, the ring will also produce an electric field in the x-y plane in the vicinity of its center. We consider this ring set-up and treat the effect of the extra electric field within perturbation theory. We calculate the total energy and show that the Landau degeneracy is now lifted. The plot of the energy as a function of $W$ now shows ``bands" (tightly spaced levels) as a consequence of the lifting of the Landau degeneracy. Like before, the plot shows that there is still a degeneracy where the bands intersect.        

\section{A charged particle moving in a linear electric field parallel to a uniform magnetic field}
The Hamiltonian for an electron of charge $-e$, mass $m_e$ and spin $\textbf{s}$ moving non-relativistically in a general electromagnetic field is given by
\beq 
H= \dfrac{1}{2\,m_e}(\textbf{p}+ e \,\textbf{A}(\textbf{x},t))^2 - e \,\phi(\textbf{x},t) -\dfrac{2\mu_e}{\hbar} \,\textbf{s}.\textbf{B}(\textbf{x},t)
\eeq{H1}
where $\textbf{p}$ is the canonical momentum, $\textbf{A}(\textbf{x},t)$ is the vector potential, $\phi(\textbf{x},t)$ is the scalar potential, $\mu_e$ is the magnetic moment of the electron and $\textbf{B}(\textbf{x},t)$ is the magnetic field. The magnetic and electric fields are given by $\textbf{B}=\bm{\nabla} \times \textbf{A}$ and $\textbf{E}=-\bm{\nabla} \phi -\dfrac{\partial \bm{A}}{\partial t}$ respectively. We now consider the case of a uniform magnetic field of magnitude $B_0$ acting in the +z direction, $\textbf{B}=B_0 \,\hat{z}$ and a linear electric field also in the z direction $\textbf{E}= \text{k}\, z \hat{z}$ where k is a positive constant. The electric field is zero on the plane $z=0$ and points in the +z direction above the plane ($z>0$) and in the -z direction below the plane ($z<0$). The magnetic and electric field therefore point in the same direction above the plane but in opposite directions below the plane. We choose the vector potential to be $A_y=x\,B_0$, $A_x=A_z=0$ whereas the scalar potential is given by $\phi(z)= -k\,z^2/2$ (chosen such that the scalar potential is zero at z=0). The choice of vector potential is of course not unique but all results are gauge invariant. Substituting the potentials into the Hamiltonian \reff{H1} yields
\beq
H= \dfrac{p_x^2}{2\,m_e}+ \dfrac{1}{2m_e} \big(p_y + e \,B_0 \,x\big)^2 + \dfrac{p_z^2}{2m_e} + \dfrac{e}{2}\,k\,z^2  -\dfrac{2\mu_e}{\hbar} \,s_z\,B_0 \,.
\eeq{H2}   
The Hamiltonian H commutes with $p_y$, $s_z$ and also $H_z=\dfrac{p_z^2}{2m_e} + \dfrac{e}{2}\,k\,z^2$ and therefore these operators share a common eigenfunction $\psi$:
\beq
H \,\psi = E \,\psi\quad;\quad H_z \psi=\varepsilon_z \psi \quad;\quad s_z \,\psi=\pm \dfrac{\hbar}{2} \,\psi \quad;\quad
 p_y \,\psi=\hbar \,k_y\,\psi \,.
\eeq{H3}
Note that $H_z$ is the Hamiltonian of a harmonic oscillator in the z-direction with angular frequency
\beq
\omega_z= \sqrt{\dfrac{ek}{m_e}}
\eeq{wz}
and energies given by  
\beq
\varepsilon_z=\Big(n_z + \dfrac{1}{2}\Big) \hbar \,\omega_z\,\quad;\quad n_z=0,1,2,3,...
\eeq{Ez}
With $H$ given by \reff{H2} and the results \reff{H3} and \reff{Ez}, the eigenvalue equation $H \,\psi=E \,\psi$ takes the form,
\beq
\Big[\dfrac{p_x^2}{2\,m_e}+ \dfrac{1}{2} m_e\,\omega_c^2\,\big(x-x_0\big)^2\Big]\psi= \Big[E-\Big(n_z+\dfrac{1}{2}\Big)
\hbar\omega_z \pm \mu_e\,B_0\Big] \psi 
\eeq{L1} 
where
\beq
\omega_c=\dfrac{e\,B_0}{m_e}
\eeq{wc}
and 
\beq
x_0=-\dfrac{\hbar \,k_y}{e\,B_0}\,.
\eeq{x0}
The operator on the left hand side of \reff{L1} corresponds to that of a harmonic oscillator moving in the x-direction with angular frequency $\omega_c$ (referred to as the cyclotron frequency) with center located at $x=x_0$. Its eigenvalues are therefore given by
\beq
E_c= \Big(n_c+ \dfrac{1}{2}\Big)\,\hbar\,\omega_c\quad;\quad n_c=0,1,2,3,...
\eeq{EC}
The above energy is degenerate since states with different values of $k_y$ have the same energy. If the magnetic field is applied to a rectangular area of sides $L_x$ and $L_y$ with periodic boundary conditions in the y direction, then $k_y$ takes on values $2\,\pi \,n_y/L_y$ where $n_y$ is a positive or negative integer. The value of $x_0$ must lie between $-L_x/2$ and $L_x/2$ so that $-\tfrac{e\,B_0\,L_xL_y}{2\,h} < n_y < \tfrac{e\,B_0\,L_xL_y}{2\,h}$. The degeneracy $D$, which is the number of possible integral values of $n_y$, is then given by the integer part of  
\beq
D= 2\,\Big(\dfrac{e\,B_0\,L_xL_y}{2\,h}\Big)=\dfrac{B_0\,A}{(h/e)}= \dfrac{\Phi}{\Phi_0}
\eeq{Degen}
where $\Phi=B_0 \,A$ is the magnetic flux through the rectangular area and $\Phi_0=\dfrac{h}{e}$ is a fundamental unit of quantum flux.   

Substituting \reff{EC} for the eigenvalue of equation \reff{L1}, one obtains  
\beq
E-\Big(n_z+\dfrac{1}{2}\Big)\hbar\,\omega_z \pm \mu_e\,B_0 = \Big(n_c+\dfrac{1}{2}\Big)\hbar\,\omega_c\,.
\eeq{L2}
The magnetic moment of the electron is given by 
\beq
\mu_e=-\dfrac{e \,\hbar\,(1+\delta)}{2\,m_e}
\eeq{mue}
where $\delta=0.0011$ is a very small radiative correction. Expressing $B_0$ in terms of $\omega_c$ via \reff{wc} we obtain
\beq
\mu_e\,B_0= - \dfrac{\hbar\, (1+\delta)}{2}\,\omega_c \,.
\eeq{mueB0}
Substituting \reff{mueB0} into \reff{L2} we obtain the exact expression for the energies:
\beq
E=\Big(n_c+\dfrac{1}{2}\pm \dfrac{(1+\delta)}{2}\Big)\hbar\,\omega_c + \Big(n_z+\dfrac{1}{2}\Big)\hbar\,\omega_z\,. 
\eeq{E}
The first term on the right hand side has a near two-fold degeneracy due to the contribution of the spin. Setting $n_c=n$ (with $n\ge 1$) and choosing the negative sign one obtains $(n-\delta/2)\hbar\,\omega_c$ whereas setting $n_c=n-1$ and choosing the positive sign yields $(n+\delta/2)\hbar\,\omega_c$. The two energies are basically equal as the ratio of the latter to the former is to first order given by  $1+\delta/n$ where $\delta/n$ is less than one in a thousand. For all practical purposes, we can therefore neglect $\delta$ and consider the two-fold degeneracy to be exact\footnote{We proceed with this approximation as the analytical results simplify considerably without much loss in numerical accuracy.}. This allows one to express the above energy in the convenient form 
\beq
E=n\,\hbar\,\omega_c+ \Big(n_z+\dfrac{1}{2}\Big)\hbar\,\omega_z \quad;\quad n,n_z=0,1,2,3,...
\eeq{E3}
where there is a two-fold degeneracy due to the spin for every $n$ except $n=0$ which is non-degenerate. The energy \reff{E3} splits neatly into two terms. The first term depends on the cyclotron frequency $\omega_c$ and is due to the uniform magnetic field only. One recognizes this term as the Landau levels with spin included. The second term depends on the angular frequency $\omega_z$ and is due to the linear electric field alone. There are no cross terms that depend on both the magnetic and electric fields. 

An important point is that if $\omega_c/\omega_z$ is a rational number, different values of $n$ and $n_z$ can yield the same energy. The degeneracy in this case will be studied in the next subsection. For all other cases -- where $\omega_c/\omega_z$ is an irrational number -- the total degeneracy is obtained by multiplying the degeneracy due to the spin with the factor $D$ given by \reff{Degen}
\beq
g_{{}_{\text{irrational}}}= 
\begin{cases}
    2D& \text{if } n\ne 0\\
    D& \text{if } n=0\,.
\end{cases}
 \eeq{g2}

\subsection{Degeneracy when $\dfrac{\omega_c}{\omega_z}$ is a rational number} 
We can express the energy \reff{E3} in the following form 
\beq
E= \Big(n \,\dfrac{\omega_c}{\omega_z}+ n_z +\dfrac{1}{2}\Big) \hbar\omega_z\,.
\eeq{E4}
We now show that different values of $n$ and $n_z$ can yield the same energy only if the ratio $\omega_c/\omega_z$ is a rational number. Let $Q= n \,\frac{\omega_c}{\omega_z}+ n_z$. Let $n$ change by an integer $i$ and $n_z$ by an integer $j$. Then the new value of $Q$, which we label $Q'$, is $Q'=(n+i) \,\frac{\omega_c}{\omega_z}+ (n_z+ j)= Q + i \,\frac{\omega_c}{\omega_z} +j$. The energy remains the same if $Q'=Q$ which implies that $\frac{\omega_c}{\omega_z}= -\frac{j}{i}$. This means that there can be a degeneracy due to different pairs $(n,n_z)$ only if $\frac{\omega_c}{\omega_z}$ is a positive rational number with $j$ and $i$ having opposite signs (i.e. if $n$ increases then $n_z$ decreases and vice versa).    

Before we proceed with analyzing the general case where $\omega_c/\omega_z$ is any rational number, it is instructive to look at the special case when the two angular frequencies are equal: $\omega_c=\omega_z=\omega$. This occurs when the magnetic field strength $B_0$ reaches the value of $\sqrt{m_e\,k/e}$. Then \reff{E4} reduces to 
\beq
E_{\omega_c=\omega_z}= \Big(n + n_z +\dfrac{1}{2}\Big)\hbar\omega= \Big(N +\dfrac{1}{2}\Big)\hbar\omega
\eeq{Ecz}
where $N=n+n_z=0,1,2,3,...$. There are many possible pairs $(n,n_z)$ that yield a given $N$ and hence a given energy. The different pairs are $(N,0),(N\m 1,1),(N\m 2, 2),...,(0,N)$. There are $N$ pairs with $n\ne 0$, each of which are two-fold degenerate due to the spin and one pair with $n=0$ which is non-degenerate. The degeneracy due to the possible $(n,n_z)$ pairs and the spin is therefore given by $2N +1$. The total degeneracy is obtained by multiplying this value with the factor $D$ given by \reff{Degen}:
\beq
g_{\omega_c=\omega_z}=(2N+1)D \,.
\eeq{g}
When the two frequencies are equal, we see that the degeneracy depends on $N$ and hence on the energy; higher energy levels are more degenerate. This is in sharp contrast to the usual Landau levels, where the degeneracy is constant for all energy levels \footnote{If one includes spin as part of the Landau levels, the ground state has half the degeneracy of the other levels i.e. the degeneracy of the Landau levels is given by \reff{g2} when spin is included.}.

We now consider the general case where $\frac{\omega_c}{\omega_z}$ is a fraction $\tfrac{I}{J}$ where $I$ and $J$ are positive integers and $J$ is the smallest possible denominator. Let $P$ be defined as 
\beq
P= n\,\tfrac{I}{J}+ n_z  \quad;\quad n,n_z=0,1,2,3,...
\eeq{P}
so that the energy \reff{E4} is given by $E= (P+\tfrac{1}{2}) \hbar\omega_z$. The degeneracy of the energy $E$ is therefore same as the degeneracy of $P$. The goal is to find the degeneracy for a given pair $(n,n_z)$. 

For a given value of $P$, there is a maximum value that $n_z$ can attain, which we label $n_{z_{max}}$. At that point, $n$ is at its minimum possible value $n_{min}$. Starting with $n_z=n_{z_{max}}$ and $n=n_{min}$, decreasing $n_z$ by $I$ and increasing $n$ by $J$ yields the same value of $P$. This procedure can be repeated $[\tfrac{n_{z_{max}}}{I}]$ times at which point $n_z$ reaches $n_{z_{min}}$, its minimum value (here $[x]$ denotes the greatest integer less than or equal to $x$).  The number of different pairs $(n,n_z)$ that yield the same value of $P$ is then $[\tfrac{n_{z_{max}}}{I}] +1$.

We would now like to express $[\tfrac{n_{z_{max}}}{I}]$ in terms of the quantum numbers $n$ and $n_z$. First note that $n_{min}<J$. We can therefore write $[\tfrac{n_{z_{max}}}{I}]=[\tfrac{n_{z_{max}}}{I}] +[\tfrac{n_{min}}{J}]$ since $[\tfrac{n_{min}}{J}]=0$. The possible values of $n$ and $n_z$ that yield the same value of $P$ are $n_z=n_{z_{max}}- \ell I$ and $n=n_{min} +\ell \,J$ where $\ell=0,1,2,3,...,[\tfrac{n_{z_{max}}}{I}]$. Substituting this into the above, we obtain that
\beq
\Big[\dfrac{n_{z_{max}}}{I}\Big]=\Big[\dfrac{n_z}{I}\Big] +\Big[\dfrac{n}{J}\Big]\,.
\eeq{nnz}

If $P$ is an integer, then $n_{z_{max}}=P$ and $n_{min}=0$. Of the $[\tfrac{n_{z_{max}}}{I}] +1$ possible pairs that yield the same value of $P$, one pair has $n=0$ and this has a spin degeneracy of unity while the remaining $[\tfrac{n_{z_{max}}}{I}]$ pairs have $n\ne 0$ and have a spin degeneracy of $2$. So the degeneracy due to both the spin and the possible $(n,n_z)$ pairs is $2\,[\tfrac{n_{z_{max}}}{I}] +1$. We must also include the degeneracy factor $D$ given by \reff{Degen}. The total degeneracy when $P$ is an integer is then given by
\beq
g_{p_{integer}}=\Big(2\Big[\dfrac{P}{I}\Big] +1\Big)\,D =\Big[2 \Big(\Big[\dfrac{n_z}{I}\Big] +\Big[\dfrac{n}{J}\Big]\Big) +1\Big]\,D \,.
\eeq{PInteger}
Note that when $I=1$, the above degeneracy is the same as that of \reff{g}.   

When $P$ is not an integer, $n_{min}\ne 0$. In that case, all the $[\tfrac{n_{z_{max}}}{I}] +1$ possible pairs have a spin degeneracy of 2.  So the degeneracy due to both the spin and the possible $(n,n_z)$ pairs is $2([\tfrac{n_{z_{max}}}{I}] +1)$. Including the degeneracy factor $D$ we obtain
\beq
g_{p\ne integer}=2 \Big(\Big[\dfrac{n_z}{I}\Big] +\Big[\dfrac{n}{J}\Big] +1\Big)\,D \,.
\eeq{PNinteger}
   
We plot below the energy \reff{E4} (in units of $\hbar \omega_z$) as a function of the ratio $W=\omega_c/\omega_z$ (you can think of $\omega_z$ as being held fixed while $\omega_c$ changes). Points of intersection occur when $W$ is rational and represent degeneracies. The red horizontal lines represent $n=0$ (the first one occurs at $n_z=0$, the second one at $n_z=1$, etc.). If there is a red line plus a number of blue lines through an intersection point, the degeneracy is given by twice the number of blue lines plus one for the red line, all multiplied by D. This corresponds to formula \reff{PInteger} for integer $P$. If there are only blue lines at an intersection point, the degeneracy is twice the number of blue lines, again all multiplied by D. This corresponds to formula \reff{PNinteger} when $P$ is rational but not an integer. Where there are no intersections, the degeneracy is D for a red line and 2D for a blue line.	 
\pagebreak 			
\begin{figure}
 \caption{\label{Energy_Levels} Energy levels (in units of $\hbar \omega_z$) as a function of the ratio $W=\omega_c/\omega_z$. Each line corresponds to an $(n,n_z)$ pair. Lines intersect where $W$ is rational and represent degeneracies. These are quite evident at ratios of unity and $2$ but can also be seen at other rational numbers.}
 \includegraphics[scale=0.6]{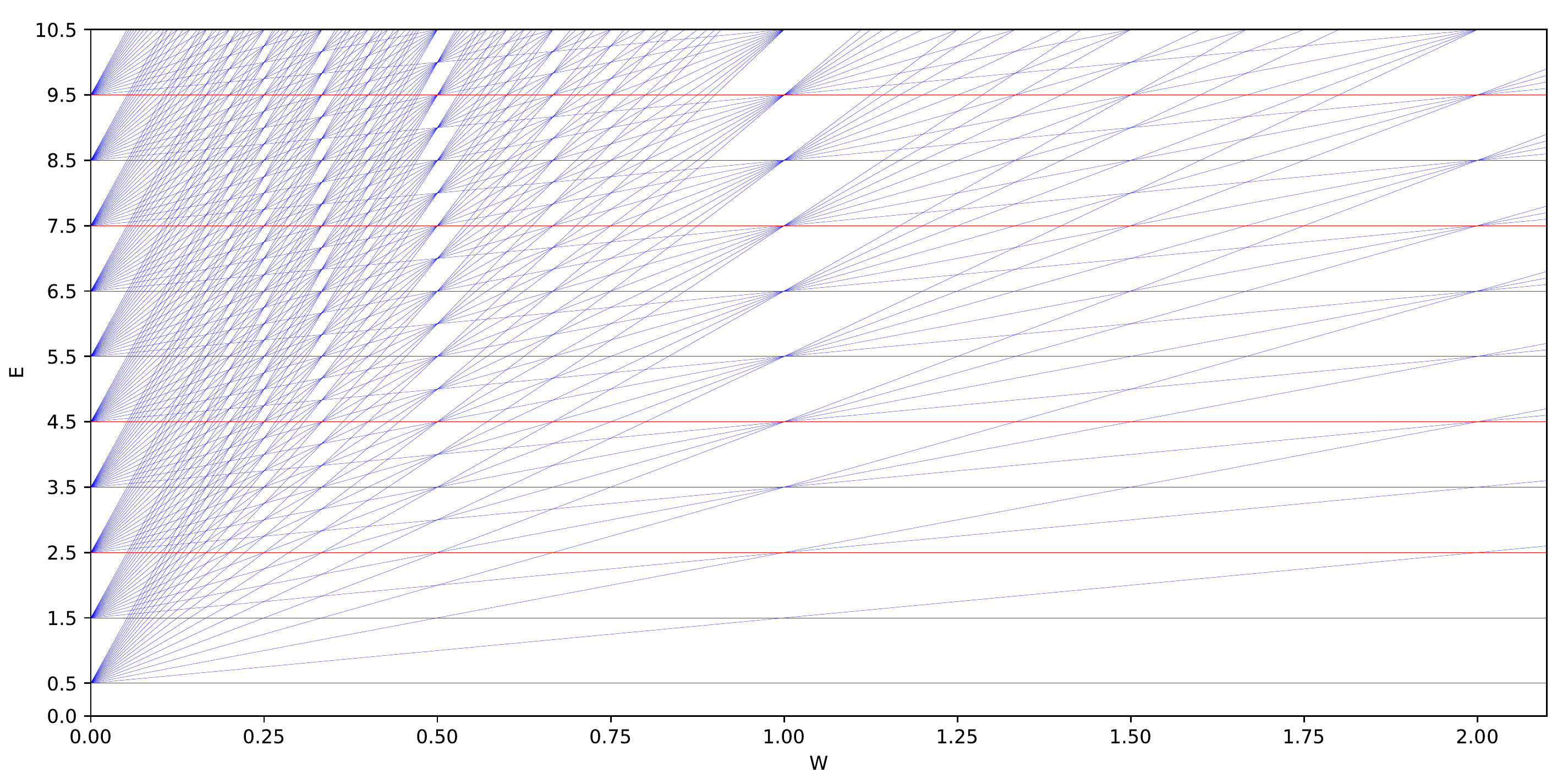}
 \end{figure}

\section{Uniformly charged ring and first-order corrections}

In section 2 we obtained an exact solution for the case of a linear electric field parallel to a constant magnetic field. How does one produce such an electric field? One can place a uniformly charged ring of radius $R$ along the x-y plane (at $z=0$). This will produce the desired linear electric field in the z direction in the vicinity of its center (and hence lead to the desired harmonic motion in the z-direction) but also produces an extra electric field in the x-y plane.  We calculate the contribution to the energy of the extra electric field using first-order perturbation theory in symmetric gauge. We will also see that the electric field along the x-y plane lifts the Landau degeneracy $D$ given by \reff{Degen}. 

\subsection{Electric field of uniformly charged ring}

In appendix A we consider a uniformly charged ring of radius $R$ and constant charge per unit length $\lambda$ placed along the x-y plane with its center at coordinates (0,0,0). The electric field at a point (x,y,z) where $|x/R|<<1$, $|y/R|<<1$ and $|z/R|<<1$ is given to first order in $x/R$, $y/R$ and $z/R$ by \reff{E2}
\beq
{\bf E} =\dfrac{\lambda}{2\epsilon_0 R^2} (z \hat{z} - \dfrac{x}{2} \hat{x}- \dfrac{y}{2} \hat{y}) = k (z \hat{z} - \dfrac{x}{2} \hat{x}- \dfrac{y}{2} \hat{y})
\eeq{electric} 
where $\epsilon_0$ is the vacuum permittivity and the constant $k$ is defined as $k \equiv \dfrac{\lambda}{2\epsilon_0 R^2}$. This is the electric field at the location of a disk with radius $r_0<<R$ and thickness $z_0<<R$ placed at the center of the ring. The scalar potential $\phi$ is then given by \reff{phi5}
\beq
\phi =-\dfrac{k}{2}(z^2- \dfrac{x^2}{2}- \dfrac{y^2}{2})\,.
\eeq{phi2}
As a check, note that the above potential produced by the ring obeys Laplace's equation $\nabla^2 \phi=0$. 

\subsection{First-order correction and symmetric gauge}
\begin{figure}[H]
\centering
		\includegraphics[scale=0.5]{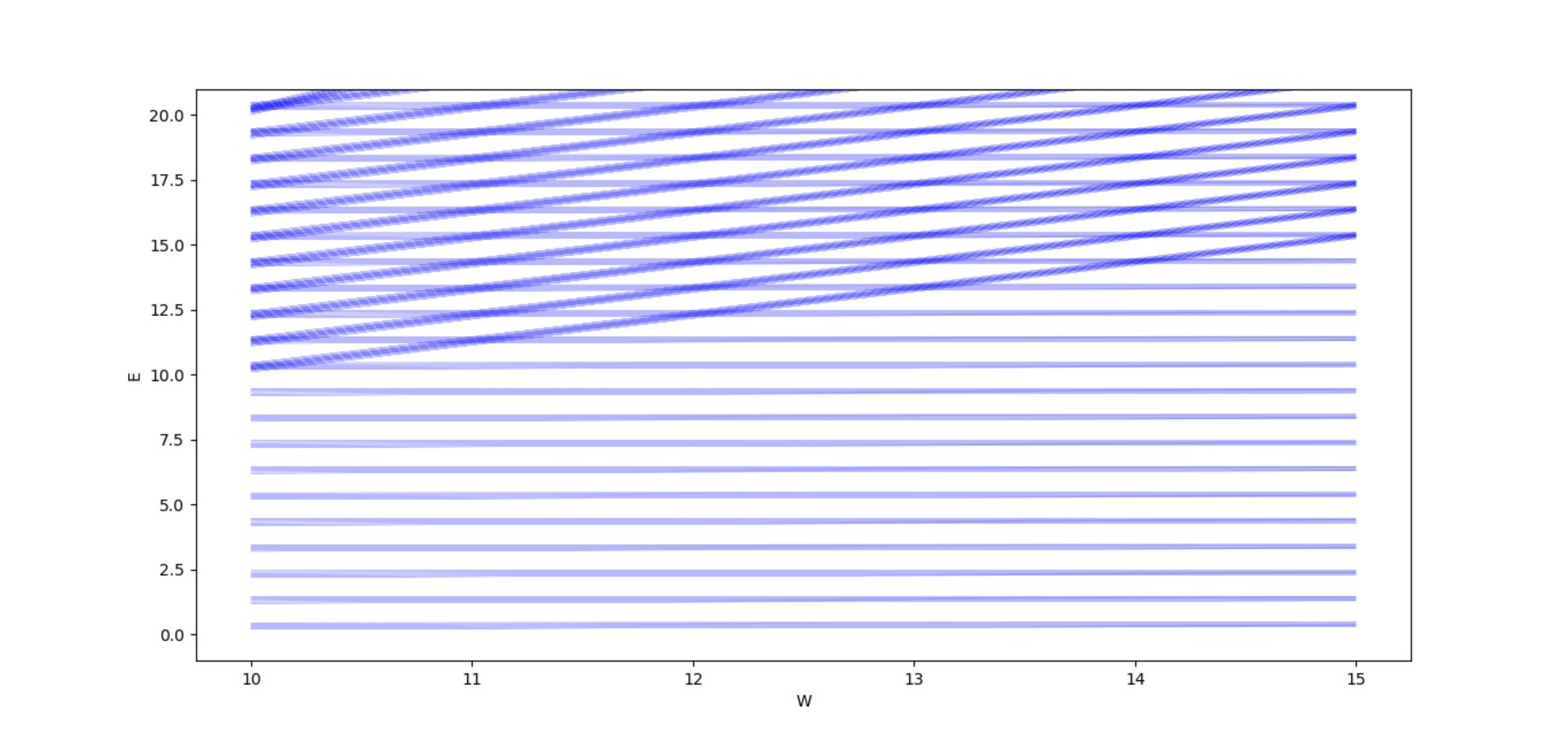}
	\caption{Energy levels (in units of $\hbar \omega_z$) as a function of $W=\omega_c/\omega_z$. There are now bands, groups of tightly spaced levels, due to the quantum number $m$ breaking the Landau degeneracy $D$. Intersection points reveal where the energy is degenerate.}
	\label{fig:Energy_ring}
\end{figure}

It is now convenient to work in symmetric gauge where the components of the vector potential ${\bf A}$ are given by $A_x= -y B_0/2$, $A_y= x B_0/2$ and $A_z=0$. Substituting this into the Hamiltonian \reff{H1} with $\phi$ given by \reff{phi2} yields
\begin{align}
H&=H_{xy} +H_z +S_z +H'=H_0 +H'
\end{align}
where 
\beq 
H_{xy}=\dfrac{1}{2\,m_e}[(p_x-e yB_0/2)^2 +(p_y +e xB_0/2)^2]  
\eeq{Hxy}
\beq
H_z= \dfrac{p_z^2}{2\,m_e}+ \dfrac{ek}{2} z^2 
\eeq{Hz}
\beq
S_z=-\dfrac{2\mu_e}{\hbar} \,s_z\,B_0
\eeq{Sz}
and 
\beq
H'=-\dfrac{ek}{4} (x^2 +y^2)\,.
\eeq{H'}
In section 2 we calculated the exact energy eigenvalues of the operator $H_0=H_{xy} +H_z +S_z$. They are given by \reff{E} i.e. identical to those of the Hamiltonian \reff{H2} except that $H_0$ is now expressed in symmetric gauge. We treat $H'$ as a perturbation and calculate its contribution to the energy using first order perturbation theory. To this end we define the following creation and annihilation operators:
\begin{align}
a&=\sqrt{\dfrac{m_e \omega_c}{2 \hbar}}\Big[\Big(\dfrac{x}{2} + \dfrac{i\,p_x}{m_e \omega_c}\Big) -i \Big(\dfrac{y}{2} + \dfrac{i\,p_y}{m_e \omega_c}\Big)\Big] \nonumber\\   
a^{\dagger}&=\sqrt{\dfrac{m_e \omega_c}{2 \hbar}}\Big[\Big(\dfrac{x}{2} - \dfrac{i\,p_x}{m_e \omega_c}\Big) +i \Big(\dfrac{y}{2} - \dfrac{i\,p_y}{m_e \omega_c}\Big)\Big] \nonumber\\ 
b&=\sqrt{\dfrac{m_e \omega_c}{2 \hbar}}\Big[\Big(\dfrac{x}{2} + \dfrac{i\,p_x}{m_e \omega_c}\Big) +i \Big(\dfrac{y}{2} + \dfrac{i\,p_y}{m_e \omega_c}\Big)\Big] \nonumber\\ 
b^{\dagger}&=\sqrt{\dfrac{m_e \omega_c}{2 \hbar}}\Big[\Big(\dfrac{x}{2} - \dfrac{i\,p_x}{m_e \omega_c}\Big) -i \Big(\dfrac{y}{2} - \dfrac{i\,p_y}{m_e \omega_c}\Big)\Big]\,.  
\end{align}
Their commutators are given by $[a, a^{\dagger}]=1$ and $[b, b^{\dagger}]=1$ (all other commutators vanish). The operator $H_{xy}= \hbar \omega_c (a^{\dagger}a +1/2)$ and the angular momentum operator $L_z= x \,p_y-y\,p_x=-\hbar(b^{\dagger}b-a^{\dagger}a)$ commute and share the eigenstate $\ket{n_c \,m}$ where $-\hbar m$ is the eigenvalue of $L_z$ and $\hbar \omega_c (n_c +1/2)$ is the eigenvalue of $H_{xy}$. We have that $a^{\dagger}a\ket{n_c\,m}=n_c\ket{n_c \,m}$ and $b^{\dagger}b\ket{n_c\, m}=(n_c+m) \ket{n_c \,m}$ with $n_c\ge 0$ and $(n_c+m)\ge 0$ (both $n_c$ and $m$ are integers). Note that in symmetric gauge, the Landau degeneracy occurs because the Landau energy $\hbar \omega_c (n_c +1/2)$ does not depend on the angular momentum quantum number $m$ i.e. there are many $m$ values for the same energy. Our goal is to evaluate the expectation value of $H'$ in the state $\ket{n_c\, m}$. We first express $x$ and $y$ in terms of the creation and annihilation operators:
\begin{align}
x&=\sqrt{\dfrac{\hbar}{2 m_e\omega_c}}\,(a+ a^{\dagger}+ b+ b^{\dagger})\nonumber\\
y&=\sqrt{\dfrac{\hbar}{2 m_e\omega_c}} \,i(a- a^{\dagger}- b+ b^{\dagger})\,.
\label{xy}
\end{align} 
Squaring the above and keeping only the terms that contribute to the expectation value yields  
\begin{align}
x^2&=\dfrac{\hbar}{2 m_e\omega_c}\,(2+2 a^{\dagger}a + 2 b^{\dagger}b)\nonumber\\
y^2&=\dfrac{\hbar}{2 m_e\omega_c}\,(2+2 a^{\dagger}a + 2 b^{\dagger}b)\,.
\label{x2}
\end{align}
Note that $x^2$ and $y^2$ make the same contribution to the expectation value. The first-order correction, denoted by $E'$,  is then given by:
\beq
E'=\bra{n_c \, m} H'\ket{n_c \, m}=-\dfrac{ek\hbar}{8 m_e \omega_c} \bra{n_c \, m}4+4 a^{\dagger}a + 4 b^{\dagger}b\ket{n_c \, m}=
-\dfrac{ek\hbar}{2 m_e \omega_c} (2n_c+m+1)\,.
\eeq{Eprime}
Expressing $k$ in terms of the frequency $\omega_z$ given by \reff{wz} we obtain
\beq
E'= -\hbar\omega_z \,\dfrac{\omega_z}{\omega_c} \Big(n_c+\dfrac{m}{2}+\dfrac{1}{2}\Big)\,.
\eeq{Eprime2}
$E'$ will be a perturbation as long as $\dfrac{\omega_z}{\omega_c}<<1$. 
Adding $E'$ to the energy \reff{E} of the unperturbed operator $H_0$ yields the total energy 
\beq
E=\Big(n_c+\dfrac{1}{2}\pm \dfrac{(1+\delta)}{2}\Big)\hbar\,\omega_c + \Big(n_z+\dfrac{1}{2}\Big)\hbar\,\omega_z -\hbar\omega_z \,\dfrac{\omega_z}{\omega_c} \Big(n_c+\dfrac{m}{2}+\dfrac{1}{2}\Big)\,.
\eeq{E5}
The Landau degeneracy $D$ given by \reff{Degen} is now lifted because the energy depends on the quantum number $m$. As before, since the radiative correction $\delta$ is much smaller than unity, we will neglect here its contribution. The $\pm$ sign in \reff{E5} leads to the following energies (denoted by $E_{+}$ and $E_{-}$ respectively):
\begin{align}
E_{+}&= \Big(n_c+1\Big)\hbar\,\omega_c + \Big(n_z+\dfrac{1}{2}\Big)\hbar\,\omega_z -\hbar\omega_z \,\dfrac{\omega_z}{\omega_c} \Big(n_c+\dfrac{m}{2}+\dfrac{1}{2}\Big)\nonumber\\
E_{-}&=n_c\,\hbar\,\omega_c + \Big(n_z+\dfrac{1}{2}\Big)\hbar\,\omega_z -\hbar\omega_z \,\dfrac{\omega_z}{\omega_c} \Big(n_c+\dfrac{m}{2}+\dfrac{1}{2}\Big)\,.
\label{Eplus}
\end{align}
It is convenient to express $E_{+}$ and $E_{-}$ in terms of the ratio $\dfrac{\omega_c}{\omega_z}$ (and its inverse):
\begin{align}
E_{+}&= \Big((n_c +1) \,\dfrac{\omega_c}{\omega_z}+ n_z+\dfrac{1}{2} -n_c\,\dfrac{\omega_z}{\omega_c}-\dfrac{(m+1)}{2}\,\dfrac{\omega_z}{\omega_c}\Big)\hbar \omega_z\nonumber\\
E_{-}&=\Big(n_c \,\dfrac{\omega_c}{\omega_z} + n_z+\dfrac{1}{2} -n_c\,\dfrac{\omega_z}{\omega_c}-\dfrac{(m+1)}{2}\,\dfrac{\omega_z}{\omega_c}\Big)\hbar \omega_z\,.
\label{Eplus2}
\end{align}
We plot the energy levels for the quantum number $m$ ranging from $0$ to $5$ in integer steps. This yields bands of tightly spaced levels and this replaces the Landau degeneracy $D$. Intersection points occur where the energy is degenerate. In figure \reff{Energy_Levels} the intersection points occurred when $W=\omega_c/\omega_z$ is rational. Below, in figure \reff{fig:Energy_ring}, they can occur at both rational and irrational values of $W$.  

\section{Conclusion}
In this work we showed that adding a linear electric field in the $z$ direction to the magnetic field of the usual Landau scenario has consequences not only for the energy levels but most importantly for the degeneracy of those levels. When the ratio of the two frequencies $W\!=\omega_c/\omega_z$ is rational, at particular energies, the degeneracy increases in contrast to the degeneracy of the Landau levels which are independent of the energy. We obtained analytical formulas for the degeneracy: equation \reff{g2} when $W$ is irrational, equation \reff{PInteger} when $P$, given by \reff{P}, is an integer and equation \reff{PNinteger} when $P$ is not an integer. We also obtained an analytical formula for the energy, equation \reff{E3}. This energy formula is valid in empty space and is determined entirely by the magnetic and electric field. 

For concreteness, we considered the electric field produced by a uniformly charged ring in the vicinity of its center. This produces the desired linear electric field in the $z$ direction but also an electric field in the x-y plane which we treat via perturbation theory. The Landau degeneracy $D$ given by \reff{Degen} is lifted and replaced by bands of tightly spaced energy levels. A plot of the energy levels as a function of $W$ reveals that there are still many points of intersection where the energy is degenerate. This time, however, the degeneracy can occur at points where $W$ is not necessarily rational.      

In the future, we would like to study the behavior of electrons in a semiconductor subject to a constant magnetic field and linear electric field. In most semiconducting materials there would be modifications to the energy formula; we would have to multiply the Zeeman term by a g-factor and replace the other energy terms by an effective energy since the effective mass of the electrons depends on the band structure. But the important finding here would persist in semiconducting materials: the degeneracy would be altered at specific ratios of $W$. There may also be reason to add a confining quadratic potential as is done in work on the Fock-Darwin system as this might model the boundaries of the material in a better fashion. If that is the case, the energy and degeneracy landscape would be quite rich because the system would now be characterized by three frequencies.     

\pagebreak
\begin{appendix}
\section{Electric field of a uniformly charged ring}
\numberwithin{equation}{section}
\setcounter{equation}{0}
Consider a uniformly charged ring of radius $R$ with constant charge per unit length $\lambda$. Let the ring be placed along the x-y plane at z=0 with its center at (0,0,0). We want to find the electric field at a point (x,y,z) where $|x/R|<<1$, $|y/R|<<1$ and $|z/R|<<1$. If a disk centered at (0,0,0) has radius $r_0<<R$ and thickness $z_0<<R$ then the conditions $|x/R|<<1$, $|y/R|<<1$ and $|z/R|<<1$ will be satisfied at the disk. 

\begin{figure}[H]
	\centering
		\includegraphics[scale=0.75]{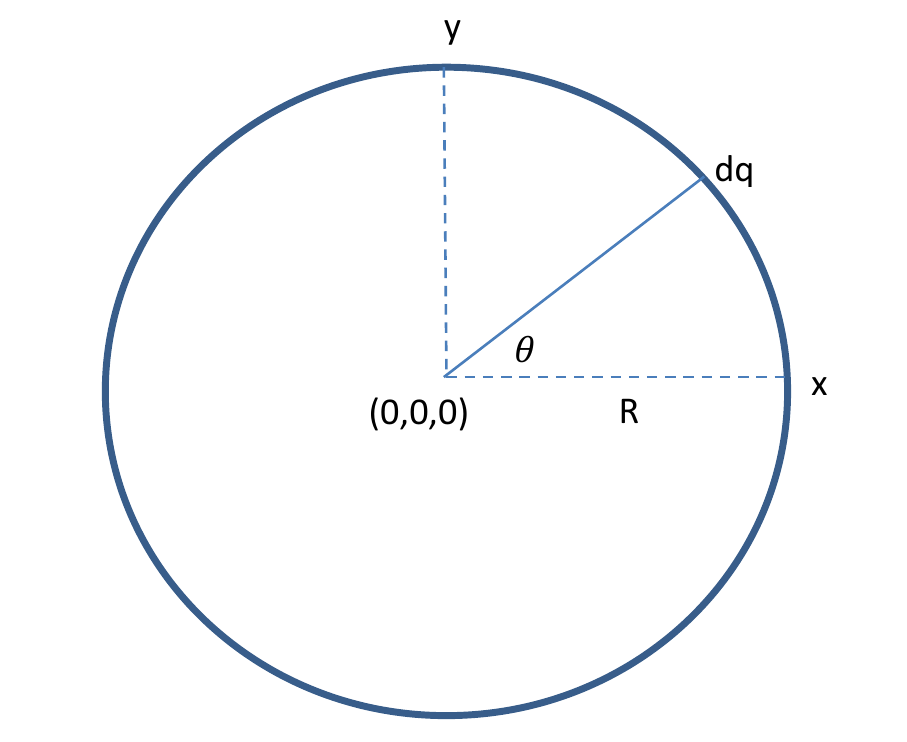}
	\caption{Uniformly charged ring on the x-y plane with center at (0,0,0).}
	\label{fig:Ring-of-Charge}
\end{figure}

Consider a differential amount of charge $dq=\lambda R d \theta$ located at coordinates $(R \cos\theta, R \sin\theta, 0)$ on the ring (see figure \ref{fig:Ring-of-Charge}). The distance $d$ to the point (x,y,z) is given by $d=[(x-R\cos\theta)^2 +(y-R \sin\theta)^2 +z^2]^{1/2}$. The electric field $d{\bf E}$ due to the charge dq is then given by
\begin{align}
d{\bf E}= \dfrac{1}{4\pi\epsilon_0}\dfrac{dq}{d^2} \hat{d}&= \dfrac{1}{4\pi\epsilon_0}\lambda R \,\dfrac{(x-R\cos\theta)\hat{x} +(y-R \sin\theta) \hat{y} +z \hat{z}}{\Big[(x-R\cos\theta)^2 +(y-R \sin\theta)^2 +z^2\Big]^{3/2}}d\theta\nonumber\\&=\dfrac{1}{4\pi\epsilon_0}\dfrac{\lambda}{R^2} \,\dfrac{(x-R\cos\theta)\hat{x} +(y-R \sin\theta) \hat{y} +z \hat{z}}{\Big[1-\dfrac{2x}{R} \cos\theta -\dfrac{2y}{R} \sin\theta + \dfrac{x^2}{R^2} +\dfrac{y^2}{R^2} + \dfrac{z^2}{R^2}\Big]^{3/2}}\,d\theta \nonumber\\&\approx\dfrac{1}{4\pi\epsilon_0}\dfrac{\lambda}{R^2} [(x-R\cos\theta)\hat{x} +(y-R \sin\theta) \hat{y} +z \hat{z}]\Big[1+\dfrac{3x}{R} \cos\theta +\dfrac{3y}{R} \sin\theta\Big] d \theta
\label{dE}
\end{align} 
where $\epsilon_0$ is the vacuum permittivity and we expanded to first order in $x/R$ and $y/R$ and neglected smaller second order terms such as $x^2/R^2$, $y^2/R^2$ and $z^2/R^2$. The components of the electric field are given by:
\begin{align}
E_x&\approx\dfrac{1}{4\pi\epsilon_0}\dfrac{\lambda}{R^2} \int_0^{2\pi} \Big(x-R\cos\theta +\dfrac{3x^2}{R} \cos\theta -3x\cos^2\theta +\dfrac{3xy}{R} \sin\theta -3 y \sin\theta \cos\theta\Big)d\theta\nonumber \\
&= -\dfrac{\lambda}{2\epsilon_0 R^2}\,\dfrac{x}{2}\nonumber\\
E_y&\approx\dfrac{1}{4\pi\epsilon_0}\dfrac{\lambda}{R^2} \int_0^{2\pi} \Big(y-R\sin\theta +\dfrac{3xy}{R} \cos\theta -3x\cos\theta \sin\theta+\dfrac{3y^2}{R} \sin\theta -3 y \sin^2\theta\Big)d\theta\label{E2}\\
&= -\dfrac{\lambda}{2\epsilon_0 R^2}\,\dfrac{y}{2}\nonumber\\
E_z&\approx\dfrac{1}{4\pi\epsilon_0}\dfrac{\lambda}{R^2} \int_0^{2\pi}\Big(z+\dfrac{3xz}{R} \cos\theta +\dfrac{3yz}{R} \sin\theta\Big)d\theta\nonumber\\
&=\dfrac{\lambda}{2\epsilon_0 R^2}\,z\,.\nonumber
\end{align}
\end{appendix}
Since ${\bf E}=-\nabla \phi$, the scalar potential $\phi$ is given by  
\beq
\phi=-\dfrac{\lambda}{2\epsilon_0 R^2} \Big(\dfrac{z^2}{2} -\dfrac{x^2}{4}-\dfrac{y^2}{4}\Big) = -\dfrac{k}{2}\Big(z^2-\dfrac{x^2}{2}-\dfrac{y^2}{2}\Big)
\eeq{phi5}
where we defined the constant $k$ to be $k\equiv \dfrac{\lambda}{2\epsilon_0 R^2} $.

%----------------------------------------------------------------------------------------
\section*{Acknowledgments}
A.E. acknowledges support from a discovery grant of the National Science and Engineering Research Council of Canada (NSERC). We thank Ren\'e C\^ot\'e of l'Universit\'e de Sherbrooke for valuable discussions and comments.

\end{document}